\documentclass[12pt]{article}
\usepackage{geometry}
\usepackage{graphicx} 
\usepackage{float}
\usepackage{nameref}  
\usepackage{amsmath}
\usepackage{authblk}
\usepackage{comment}
\usepackage{multirow}
\usepackage{multicol}
\usepackage{xcolor}
\usepackage{url}
\usepackage[T1]{fontenc}

\usepackage{rotating}  
\usepackage{booktabs}  
\usepackage{multirow}  

\title{Complementary Thermodynamic Mechanisms of Boron and Carbon Segregation at Grain Boundaries in Nickel Alloys}

\author[1,2]{Tyler D. Dole\v{z}al\footnote{corresponding authors: dolez120@mit.edu; rodrigof@mit.edu; liju@mit.edu}}
\author[1]{Rodrigo Freitas$^*$}
\author[1,3]{Ju Li$^*$}
\affil[1]{Department of Materials Science and Engineering, Massachusetts Institute of Technology, Cambridge, MA, USA}
\affil[2]{Department of Engineering Physics, Air Force Institute of Technology, Wright-Patterson Air Force Base, OH, USA}
\affil[3]{Department of Nuclear Science and Engineering, Massachusetts Institute of Technology, Cambridge, MA, USA}

\begin{document}

\maketitle

\begin{abstract}
\noindent
Grain boundary stabilization by light interstitials is central to the performance of Ni-based superalloys, yet the thermodynamic mechanisms governing their interactions with substitutional chemistry remain poorly resolved. Here, we use hybrid Monte Carlo molecular dynamics simulations to quantify how boron and carbon modify the thermodynamic, structural, and chemical ordering of grain boundaries in Ni–Cr alloys. By analyzing interfacial state variables, site-resolved segregation spectra, local chemical ordering, and structural evolution, we show that boron and carbon stabilize grain boundaries through complementary pathways. Carbon drives saturation-controlled stabilization by recruiting Cr and conditioning boundary chemistry, while suppressing temperature-driven structural transformations of the boundary. In contrast, boron stabilizes grain boundaries through a selective mechanism that lowers the interfacial grand potential via localized ordering while permitting gradual structural evolution. These effects arise from coupled interactions between interstitial segregation and Cr redistribution, which together regulate site accessibility, chemical competition, and the range of accessible interfacial states. This work provides a thermodynamic framework for grain boundary engineering and suggests design principles for leveraging interstitial–substitutional interactions in alloys.
\end{abstract}

\section*{Introduction}

Light interstitial elements boron and carbon are added to nickel-based superalloys to strengthen grain boundaries and improve high-temperature mechanical performance~\cite{simsSuperalloysIIHighTemperature1987,reedSuperalloysFundamentalsApplications2006,goldschmidtInterstitialAlloys1967}. However, their thermodynamic behavior at grain boundaries remains incompletely understood, leading to contradictory experimental observations~\cite{cadelAtomProbeTomography2002a,yanEffectBoronAdditions2008,zhangNanosizeBorideParticles2008,ojoAnalyticalElectronMicroscopy2008,hongMechanismSerratedGrain2012, alamSegregationNiBasedSuperalloy2012}. Boron exhibits dual character: at low activities it segregates elementally to grain boundaries and enhances cohesion, while at higher activities it interacts with elements like Cr to form borides~\cite{kontisEffectBoronGrain2016,duPrecipitationEvolutionGrain2017a,quEffectsBoronAddition2024,tekogluSuperiorHightemperatureMechanical2024d}. Carbon shows similar complexity, forming discrete MC carbides and continuous Cr-based M$_{23}$C$_6$ films depending on processing conditions~\cite{kontisEffectBoronGrain2016,duPrecipitationEvolutionGrain2017a,wangInsightLowCycle2023a,zhangSynergyPhaseMC2024}. The thermodynamic conditions governing the transition between elemental segregation and compound formation, and the competitive interactions between boron and carbon at grain boundaries, remain unmapped. This gap limits rational grain boundary chemistry design in high-temperature structural alloys, requiring precise interfacial control for creep resistance and long-term stability.

Resolving these contradictions requires treating grain boundaries as distinct thermodynamic phases with their own stability criteria rather than static structural defects. Early theoretical models predicted that solute segregation could induce two-dimensional phase transformations within grain boundaries, creating interfacial analogues of bulk phase transitions~\cite{hartTwodimensionalPhaseTransformation1968}. Subsequent atomistic simulations confirmed that such transformations can occur well below the melting temperature, revealing that grain boundaries possess independent thermodynamic states governed by temperature, composition, and local structure~\cite{carrionEvidenceGrainBoundary1983,frolovStructuralPhaseTransformations2013}. The complexion framework~\cite{dillonComplexionNewConcept2007a} formalized this picture by extending Gibbsian thermodynamics to internal interfaces, providing a foundation for constructing interfacial phase diagrams. Subsequent simulations demonstrated that variations in solute concentration can trigger first-order transformations between distinct boundary structures~\cite{frolovSegregationinducedPhaseTransformations2015}, while direct imaging experiments confirmed the coexistence of multiple grain boundary phases in elemental metals~\cite{meinersObservationsGrainboundaryPhase2020}. These developments established that grain boundaries undergo structural and chemical phase transitions governed by the same thermodynamic principles as bulk systems, creating a framework within which interstitial segregation behavior can be systematically analyzed.

Within this framework, segregation-driven stabilization has been extensively studied for substitutional solutes. Theoretical analyses showed that solute adsorption can reduce the grain boundary energy $\gamma$ to zero, eliminating the driving force for grain growth and enabling thermodynamic stabilization of nanocrystalline structures~\cite{kirchheimGrainCoarseningInhibited2002,liuNanoscaleGrainGrowth2004,shvindlermanUnexploredTopicsPotentials2006}. Atomistic and Monte Carlo simulations linked segregation and short-range ordering to substantial reductions in interfacial energy~\cite{detorGrainBoundarySegregation2007}, while thermodynamic and statistical models established the conditions for nanocrystalline stability through segregation enthalpy~\cite{chookajornThermodynamicsStableNanocrystalline2014,murdochStabilityBinaryNanocrystalline2013} and identified solute combinations capable of achieving this stabilization through favorable segregation energetics~\cite{darlingMitigatingGrainGrowth2014}. Statistical mechanics treatments further revealed that grain boundaries exhibit a spectrum of metastable states forming a continuous energy band~\cite{hanGrainboundaryMetastabilityIts2016}, while machine-learning and evolutionary-search methods uncovered extensive polymorphism across the misorientation space of elemental metals~\cite{zhuPredictingPhaseBehavior2018}. More recent theoretical work unified kinetic and thermodynamic descriptions of interface stability~\cite{husseinModelThermodynamicStabilization2024,liLinePlanarDefects2025}, while experimental studies confirmed that solute segregation can directly trigger grain boundary structural transformations, linking chemical decoration to enhanced cohesion and thermal stability~\cite{zhouBoronTriggersGrain2025}.

However, these advances have focused almost exclusively on substitutional solutes in elemental or simple binary systems, leaving the role of light interstitials in substitutionally complex alloys critically underexplored. Unlike substitutional solutes, interstitials occupy distinct crystallographic sites and exhibit strong short-range interactions with specific substitutional species. In Ni-based superalloys, boron and carbon segregation is mediated by Cr, the primary substitutional alloying element that governs carbide and boride phase formation~\cite{kontisEffectBoronGrain2016,duPrecipitationEvolutionGrain2017a}. Whether boron and carbon compete or cooperate at grain boundaries, and how Cr modifies their segregation thermodynamics, has not been systematically established. Prior computational studies have examined boron or carbon segregation in pure Ni or Fe matrices, but these cannot capture the concentration-dependent competition and synergy that emerge when multiple interstitial species interact with substitutional chemistry. Furthermore, existing experimental observations reveal dual regimes (elemental segregation versus compound formation) without establishing the thermodynamic criteria that govern transitions between these states or the conditions under which cooperative stabilization becomes favorable. Light interstitial elements are known to alter cohesion, diffusivity, and high-temperature performance in both steels~\cite{banerjiBoronSteelProceedings1980,dongDualRoleBoron2025} and Ni-based superalloys~\cite{reedSuperalloysFundamentalsApplications2006,xiaoEffectBoronCarbon2005}, yet the thermodynamic mechanisms underlying these effects, and their dependence on substitutional alloying, remain unresolved.

Here we use hybrid Monte Carlo (MC) and molecular dynamics (MD) simulations to quantify how light interstitials modify the thermodynamic state of grain boundaries in Ni--Cr alloys. By varying temperature (300--1200~K) and global Cr content (10--25~at.\%) at fixed, experimentally relevant chemical potentials for boron and carbon, we track the evolution of interfacial state variables across both ordered ($\Sigma5\,[001]\,(210)$ symmetric tilt) and disordered (high-angle) grain boundaries. This approach resolves how interstitial identity, substitutional chemistry, boundary character, and local chemical ordering jointly govern interfacial stability, segregation behavior, and structural response, including temperature- and composition-dependent transitions between distinct grain boundary configurations.

We demonstrate that boron and carbon stabilize grain boundaries through distinct thermodynamic pathways mediated by substitutional Cr, while simultaneously reshaping the accessible structural states of the boundary. Boron operates through selective stabilization, producing consistently lower interfacial grand potential excess despite modest enrichment by localizing into energetically favorable interfacial environments, with the expression of this selectivity depending on boundary character. Carbon stabilizes grain boundaries through a saturation-driven mechanism, extensively enriching the interface while co-segregating with Cr to chemically condition boundaries with high segregant capacity and suppressing structural evolution over a broad temperature range.

These mechanisms are not inherently competitive. Carbon establishes Cr-rich interfacial environments through saturation-driven co-segregation, while boron remains thermodynamically driven to segregate to these chemically enriched boundary states. This behavior is supported by targeted co-doping simulations in which boron is introduced into pre-equilibrated carbon-conditioned grain boundaries, indicating that boron can incorporate into carbon- and Cr-rich interfaces without displacing existing segregation. 

Together, these results reveal that interfacial thermodynamics, segregation, and structure are intrinsically coupled, with interstitial species governing not only chemical enrichment but also the stability and evolution of grain boundary configurations. This framework provides a mechanistic thermodynamic basis for interpreting experimental observations of boron-mediated modification of carbide formation under certain processing conditions, offers a consistent interpretation for the widespread use of boron--carbon-containing alloys~\cite{kontisEffectBoronGrain2016,duPrecipitationEvolutionGrain2017a,xiaoEffectBoronCarbon2005,garosshenEffectsZrStructure1987}, and suggests thermodynamic design principles for stabilizing grain boundaries in Ni-based alloys.


\section*{Results}

\subsection*{Interstitial Effects on Grain Boundary Structural Transformations}

\begin{figure}[!ht]
    \centering
    \includegraphics[width=\linewidth]{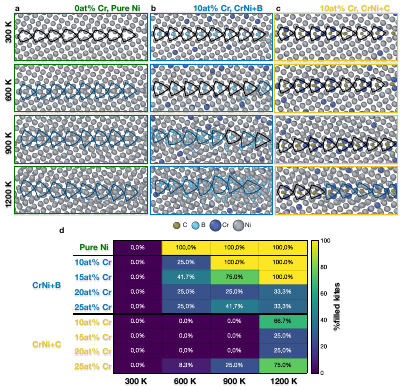}\\
    \caption{\textbf{Temperature-dependent grain boundary structural evolution and interstitial effects on kite filling.} (\textbf{a}-\textbf{c}) Atomic configurations illustrating the evolution of the $\Sigma5\,[001]\,(210)$ grain boundary structure as a function of temperature for (\textbf{a}) pure Ni (undoped and 0 at.\% Cr), (\textbf{b}) 10 at.\% Cr with B interstitial doping, and (\textbf{c}) 10 at.\% Cr with C interstitial doping. Black outlines denote open kites while blue outlines denote filled kites. (\textbf{d}) Heat map summarizing the percent of filled kite units as a function of temperature and Cr concentration for pure Ni, CrNi with boron, and CrNi with carbon systems. The data reveal a sharp temperature-driven transition in pure Ni, a composition- and temperature-dependent gradual transition in boron-doped systems, and a pronounced stabilization of open kite structures in carbon-doped systems, with significant filling occurring only at the highest temperatures and Cr concentrations. Atomic renderings visualized in OVITO~\cite{stukowskiVisualizationAnalysisAtomistic2009a}.}
    \label{fig:complexions}
\end{figure}

The temperature-dependent structural evolution of the $\Sigma5\,[001]\,(210)$ (S5) grain boundary is summarized in Fig.~\ref{fig:complexions}, which combines representative atomic configurations with a quantitative classification of kite unit filling (Table~\ref{tab:kites}). Here, we distinguish between open kite units, in which the central kite site remains part of the interstitial sublattice, and filled kite units, in which a substitutional atom (Ni or Cr) occupies this site following a local lattice reconstruction (blue outlines in Fig.~\ref{fig:complexions}a). Importantly, the open configuration is not vacant; rather, it is accessible to interstitial species, whose occupation plays a critical role in stabilizing the boundary structure. Kite filling fractions are evaluated by counting the 12 projected kite repeat units along the grain boundary plane in the $x$-direction, such that each counted unit represents a column of kite motifs extending through the simulation cell thickness ($z$-direction).

As shown in Fig.~\ref{fig:complexions}a, the pure Ni grain boundary undergoes a sharp structural transition between 300~K and 600~K, characterized by the conversion of open kite units into fully filled configurations. This transition is abrupt and complete, with all kite units adopting the filled configuration at temperatures $\geq 600$~K, consistent with a well-defined temperature-driven structural transformation associated with lattice relaxation into a lower-energy configuration.

The introduction of interstitial dopants significantly modifies this behavior. In boron-doped systems (Fig.~\ref{fig:complexions}b), the transition from open to filled kite units becomes both temperature- and composition-dependent, proceeding gradually with increasing temperature and Cr concentration. As shown in Fig.~\ref{fig:complexions}d, intermediate states containing a mixture of open and filled kite units emerge over a broad range of conditions, indicating a progressive structural evolution rather than a discrete transition. This behavior arises from the occupation of the central kite site by boron, which inhibits the lattice reconstruction required for substitutional atoms to collapse into the filled configuration. Notably, cooperative segregation between boron and Cr further enhances this effect, reducing the fraction of filled kite units from fully filled in Ni-rich boundaries to approximately 33\% in Cr-containing systems at 1200~K.

In contrast, carbon-doped systems exhibit a markedly different response. As shown in Fig.~\ref{fig:complexions}c,d, the open kite configuration is strongly stabilized across all compositions up to 900~K, with significant kite filling occurring only at the highest temperatures and/or Cr concentrations. This suppression of kite filling reflects the strong tendency for carbon to co-segregate with Cr, rapidly saturating the interstitial sublattice and stabilizing the open kite configuration. Consequently, carbon acts to suppress the temperature-driven structural transition observed in the pure and boron-doped systems across a wide range of conditions. At higher Cr concentrations (25~at.\%), however, a gradual transformation toward filled kite units is observed, reaching approximately 75\% filled at 1200~K. This suggests that at sufficiently high solute enrichment, the filled configuration becomes energetically favorable despite the presence of interstitial species.

This behavior reflects competition between interstitial site occupancy and substitutional lattice collapse, highlighting the role of interstitial–substitutional coupling in governing grain boundary complexion transitions.

\begin{table}[!ht]
\centering
\resizebox{\textwidth}{!}{%
\begin{tabular}{c c| cc| cc| cc| cc}
\toprule\toprule
\multirow{2}{*}{\textbf{System}} & \multirow{2}{*}{$\mathbf{x_{\mathrm{\textbf{Cr}}}}$ \textbf{(at.\%)}} 
& \multicolumn{2}{c|}{\textbf{300 K}} 
& \multicolumn{2}{c|}{\textbf{600 K}} 
& \multicolumn{2}{c|}{\textbf{900 K}} 
& \multicolumn{2}{c}{\textbf{1200 K}} \\
& & \textbf{Open} & \textbf{Filled} & \textbf{Open} & \textbf{Filled} & \textbf{Open} & \textbf{Filled} & \textbf{Open} & \textbf{Filled} \\
\midrule

\multirow{1}{*}{\textbf{Pure Ni}}
& 0
& 12 & 0
& 0 & 12
& 0 & 12
& 0 & 12 \\

\midrule

\multirow{4}{*}{\textbf{Boron-doped}}
& 10
& 12 & 0
& 9 & 3
& 0 & 12
& 0 & 12 \\
& 15
& 12 & 0
& 7 & 5
& 3 & 9
& 0 & 12 \\
& 20
& 12 & 0
& 9 & 3
& 9 & 3
& 8 & 4 \\
& 25
& 12 & 0
& 9 & 3
& 7 & 5
& 8 & 4 \\

\midrule

\multirow{4}{*}{\textbf{Carbon-doped}}
& 10
& 12 & 0
& 12 & 0
& 12 & 0
& 4 & 8 \\
& 15
& 12 & 0
& 12 & 0
& 12 & 0
& 9 & 3 \\
& 20
& 12 & 0
& 12 & 0
& 12 & 0
& 9 & 3 \\
& 25
& 12 & 0
& 11 & 1
& 9 & 3
& 3 & 9 \\

\bottomrule\bottomrule
\end{tabular}}
\caption{Temperature-dependent evolution of kite structures for pure and interstitial-decorated S5 grain boundaries. Entries report the number of open and filled kite units (out of 12 total units).}
\label{tab:kites}
\end{table}

\subsection*{Interfacial state variables and thermodynamic trends}

\begin{figure}[!ht]
    \centering
    \includegraphics[width=\linewidth]{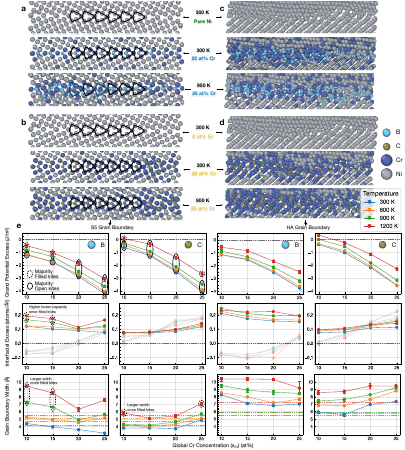}
    \caption{\textbf{Atomic renderings and interfacial state variable analysis for chemically decorated grain boundaries.} Equilibrated structures for (\textbf{a},\textbf{b}) $\Sigma5\,[001]\,(210)$ (S5) and (\textbf{c},\textbf{d}) high-angle (HA) grain boundaries. A pure Ni reference structure is included at the top of each panel for comparison with chemically decorated configurations. Black outlines denote open kites while blue outlines denote filled kites. (\textbf{e}) Grand potential excess ($\tilde{\Omega}_{\mathrm{ex}}$), interfacial excess ($\Gamma_i$), and boundary width ($w$) as functions of temperature and global Cr content. Semi-transparent curves show $\Gamma_{\mathrm{Cr}}$. Dashed lines in the $w$ plots denote the average widths of the corresponding pure Ni boundaries. All results are obtained at fixed interstitial chemical potentials corresponding to dilute boron and carbon concentrations (0.01 and 0.1 at.\%).}
    \label{fig:state_variables}
\end{figure}

To quantify the thermodynamic and structural signatures of the chemo-structural states identified in the previous section, we next examine the evolution of key grain boundary state variables across composition and temperature. While Fig.~\ref{fig:complexions} establishes that interstitial chemistry governs the structural state of the S5 grain boundary, a quantitative description is required to compare the stability and structural characteristics of these configurations across different boundary types.

Fig.~\ref{fig:state_variables} provides both a visual reference for the equilibrated grain boundary structures and a consolidated view of the interfacial state variables used throughout the remainder of the analysis. Fig.~\ref{fig:state_variables}a–d show representative post-equilibration configurations of the S5 and custom high-angle (HA) grain boundaries following hybrid semi-grand/grand canonical Monte Carlo sampling and subsequent canonical molecular dynamics relaxation. These atomic renderings are intended to orient the reader to the structural character of the two boundary types and to illustrate the spatial distribution of boron and carbon interstitials after equilibration. Consistent with the observations in Fig.~\ref{fig:complexions}, the S5 interface retains a recognizable periodic motif along the boundary plane, whereas the HA boundary lacks well-defined structural units and instead exhibits a broader, more heterogeneous interfacial region.

For the S5 boundary, the underlying kite framework remains clearly traceable across all conditions considered, with interstitial occupancy evolving within a largely preserved structural motif. As established in Fig.~\ref{fig:complexions}, increasing temperature and Cr concentration promote the progressive filling of kite units, accompanied by increasing local chemical and structural disorder. However, this disorder remains spatially confined to the immediate boundary region rather than propagating into the adjoining grains. In boron-decorated S5 boundaries, an ordered, repeating interfacial motif is observed at 300~K for the 25~at.\%~Cr composition, whereas at elevated temperatures the boundary exhibits partially and fully filled kite configurations. In contrast, carbon decoration promotes near-complete Cr occupation of the S5 boundary core while stabilizing open kite configurations over a broader temperature range, resulting in a compositionally saturated yet structurally persistent interfacial state.

Unlike the S5 boundary, the high-angle (HA) boundary exhibits no analogous repeating structural unit. Instead, increasing Cr content and temperature lead to a progressive broadening of the disordered interfacial region, accompanied by more diffuse interstitial accommodation. In the absence of discrete structural motifs, structural evolution in the HA boundary is reflected in changes to its width and chemical distribution. These qualitative differences highlight fundamentally distinct responses of ordered and disordered grain boundaries to chemical decoration, motivating the quantitative analysis of interfacial stability, segregation, and structural state introduced below.

To quantify these chemo-structural differences and track their evolution with composition and temperature, we next examine three interfacial state variables as functions of the global Cr concentration ($x_{\mathrm{Cr}}$) and temperature ($T$). As defined in the methodology, these variables comprise the grand potential excess per unit area ($\tilde{\Omega}_{\mathrm{ex}}$), which quantifies the thermodynamic stability of the grain boundary relative to a temperature-dependent reference pure Ni grain boundary; the Gibbsian interfacial excess ($\Gamma_i$), which characterizes solute and interstitial segregation; and the grain boundary width ($w$), which serves as a proxy for the structural extent of the interfacial region. 

In evaluating $\tilde{\Omega}_{\mathrm{ex}}$, the reference state is obtained from post-MD relaxed configurations of the undoped pure Ni grain boundary, such that the reference reflects the equilibrium complexion of the pure system at each temperature (open kite configurations at 300~K and filled kite configurations at elevated temperatures). This ensures that comparisons of interfacial stability are made relative to the appropriate structural state of the boundary under each thermodynamic condition.

The sign of $\tilde{\Omega}_{\mathrm{ex}}$ therefore reflects the relative thermodynamic stability of a given grain boundary configuration with respect to this reference state: negative values ($\tilde{\Omega}_{\mathrm{ex}} < 0$) indicate enhanced stability compared to the pure Ni boundary, while positive values ($\tilde{\Omega}_{\mathrm{ex}} > 0$) indicate reduced stability. Importantly, because the reference state is itself a grain boundary rather than the bulk crystal, the sign of $\tilde{\Omega}_{\mathrm{ex}}$ does not correspond to an absolute driving force for grain boundary formation, but instead provides a measure of relative stability between interfacial states.

A comprehensive compilation of these quantities for both the S5 and HA grain boundaries is presented in Fig.~\ref{fig:state_variables}e, with numerical values provided in Table~S4 of the Supplementary Information.

Across both boundary types, boron-decorated configurations consistently exhibit lower values of $\tilde{\Omega}_{\mathrm{ex}}$ than their carbon-decorated counterparts over the full range of Cr concentrations and temperatures examined. This persistent ordering indicates that, under identical thermodynamic conditions, boron decoration produces a more favorable interfacial state than carbon decoration. Crossover behavior may emerge at higher carbon activities, lower boron activities, or in the presence of additional alloying elements that further strengthen C–Cr interactions.

Increasing temperature leads to a systematic increase in the grand potential excess for both interstitial species, with $\tilde{\Omega}_{\mathrm{ex}}$ becoming less negative across the temperature range examined. This trend reflects a reduction in thermodynamic favorability relative to the corresponding pure Ni grain boundary reference states.

Further inspection reveals a clear coupling between interstitial stabilization and Cr segregation at the grain boundary. Across both boundary types, less negative and, in some cases, positive values of $\tilde{\Omega}_{\mathrm{ex}}$ coincide with a reduction in $\Gamma_\mathrm{Cr}$, indicating that Cr decoration plays a critical role in driving thermodynamic stabilization alongside interstitial segregation. Configurations with diminished Cr excess are systematically less stable for both boron and carbon, independent of boundary structure.

Boron-decorated grain boundaries maintain negative values of $\tilde{\Omega}_{\mathrm{ex}}$ across all Cr concentrations, temperatures, and boundary types examined, demonstrating that boron segregation remains thermodynamically favorable in both Ni-rich and Cr-decorated environments. This stability persists even in cases where $\Gamma_\mathrm{Cr}$ becomes slightly negative, indicating that boron does not rely on Cr co-segregation for interfacial stabilization.

In contrast, carbon exhibits conditions under which the grand potential excess becomes weakly positive. At $T = 1200~\mathrm{K}$ and $x_{\mathrm{Cr}} = 10~\mathrm{at.}\%$, carbon-decorated configurations show a positive $\tilde{\Omega}_{\mathrm{ex}}$ in both boundary types, correlated with minimal or negative $\Gamma_\mathrm{Cr}$. This behavior is further accentuated in the HA boundary, where carbon stabilization deteriorates in conjunction with Cr depletion from the interface, indicating that carbon-driven stabilization is strongly dependent on the availability of Cr-rich environments.

Finally, the thermodynamic response depends on grain boundary structure. Both boron- and carbon-decorated configurations exhibit more negative grand potential excess values in the S5 grain boundary than in the HA boundary. For boron, this enhanced stabilization in the S5 boundary coincides with lower $\Gamma_\mathrm{Cr}$, whereas for carbon it persists despite elevated excesses of both carbon and Cr relative to the S5 structure. This contrast indicates that the thermodynamic impact of interstitial and solute decoration depends not only on interfacial composition but also on the underlying structural order of the boundary.

The Gibbsian interfacial excess values further elucidate how interstitial segregation reshapes local boundary chemistry and couples to the observed complexion transitions. Both boron and carbon exhibit positive interfacial excess across all conditions studied ($\Gamma_I$), confirming preferential segregation to the grain boundary relative to the bulk. In contrast, the associated Cr interfacial excess ($\Gamma_{\mathrm{Cr}}$) exhibits markedly different responses depending on the identity of the segregating interstitial, revealing distinct co-segregation behaviors.

In boron-decorated systems, increasing boron enrichment is accompanied by strongly negative Cr interfacial excess, indicating substantial Cr depletion from the boundary core. This depletion persists across both boundary types and is most pronounced for the HA boundary, which does not exhibit positive $\Gamma_\mathrm{Cr}$ until the highest reservoir concentration of 25~at.\%. These trends indicate that Cr enrichment is energetically unfavorable in the presence of excess boron, even at elevated global Cr content. Similar boron-driven Cr depletion has been reported in density functional theory studies of B and Cr segregation at grain boundaries in austenitic Fe-based alloys, indicating that this behavior is robust across face-centered cubic Ni- and Fe-based systems~\cite{yanEffectCoSegregationCCr2023}.

In contrast, carbon-decorated grain boundaries show a steady increase in Cr interfacial excess with increasing $x_{\mathrm{Cr}}$, with the magnitude of Cr enrichment substantially exceeding that observed in boron-containing systems. This behavior indicates strong cooperative segregation between carbon and chromium, with carbon acting as an effective driver of Cr accumulation at the grain boundary under otherwise identical thermodynamic conditions.

The interfacial excess of the interstitial species themselves further highlights distinct segregation mechanisms. Carbon interfacial excess remains relatively constrained across temperature, reflecting a saturation-driven process in which available interstitial sites are rapidly occupied. In contrast, boron interfacial excess exhibits a stronger dependence on both temperature and composition, with $\Gamma_{\mathrm{B}}$ increasing systematically with temperature but decreasing with increasing bulk Cr concentration. These trends indicate that boron segregation is more sensitive to the evolving thermodynamic and chemical environment of the boundary.

Importantly, the evolution of interfacial excess is strongly coupled to the underlying grain boundary structure. In boron-decorated systems, increases in $\Gamma_{\mathrm{B}}$ coincide with the transition toward majority filled kite configurations, indicating an enhanced capacity for boron segregation following structural transformation. This suggests that the filled kite configuration provides a greater number or more favorable distribution of interstitial-accessible sites. In contrast, carbon-decorated systems exhibit relatively stable interfacial excess across temperature until sufficiently high Cr content and temperature drive partial structural transformation, at which point an increase in $\Gamma_{\mathrm{C}}$ is observed.

A similar distinction is observed in the HA grain boundary. Boron-decorated HA boundaries exhibit a broader variation in interfacial excess with temperature, consistent with a boundary structure that undergoes more continuous reorganization. In contrast, carbon-decorated HA boundaries show relatively constrained interfacial excess values, indicating a more stable structural configuration once interstitial sites are saturated through strong C–Cr co-segregation.

The absolute capacity for interstitial segregation further depends on boundary character. Across all conditions studied, the HA boundary accommodates higher interfacial excesses than the S5 boundary for both boron and carbon. However, the relative dominance of each interstitial differs: in the S5 boundary, carbon exceeds boron at high Cr concentration for most temperatures, whereas in the HA boundary, boron maintains a higher interfacial excess across the full thermodynamic range.

These results demonstrate that interfacial excess and grain boundary structure are intrinsically coupled. Interstitial segregation governs the stability of specific atomic configurations, while the resulting structural state of the boundary determines its capacity for further solute incorporation.

The segregation trends described above are accompanied by systematic changes in the effective grain boundary width, providing a structural manifestation of the coupled chemical and configurational evolution. Across all temperatures, boron-decorated boundaries remain broader than their carbon-decorated counterparts, while carbon-containing boundaries exhibit narrower interfacial regions under otherwise identical conditions. 

These differences reflect the distinct segregation mechanisms of the two interstitial species. In boron-decorated systems, the temperature- and composition-dependent evolution of $\Gamma_{\mathrm{B}}$, coupled with progressive structural reorganization, promotes a more diffuse and adaptable boundary structure. In contrast, the saturation-driven segregation of carbon and its strong co-segregation with chromium stabilize a more compact and structurally constrained interface.

The dependence on bulk Cr concentration further highlights this distinction. In boron-decorated systems, increasing $x_{\mathrm{Cr}}$ leads to progressive boundary narrowing, consistent with reduced boron interfacial excess and diminished structural disorder. Conversely, carbon-decorated boundaries widen with increasing Cr content, reflecting enhanced Cr–C co-segregation and the associated expansion of the interfacial region as solute enrichment increases.

For both the S5 and HA boundaries, the effective boundary width in boron-decorated systems correlates primarily with $\Gamma_\mathrm{B}$ and shows only a weak dependence on the accompanying $\Gamma_\mathrm{Cr}$. Increasing boron enrichment leads to a more spatially confined interfacial region, consistent with enhanced B–Cr ordering within the boundary. However, with increasing temperature, the boundary width exhibits a systematic increase, particularly in compositions where kite filling is observed in the S5 boundary. This behavior indicates that the onset of structural reconfiguration is accompanied by a transition to a more spatially extended interfacial state.

In contrast, the boundary width in carbon-decorated systems reflects the combined influence of $\Gamma_\mathrm{C}$ and $\Gamma_\mathrm{Cr}$ for both boundary types, with increasing carbon enrichment generally accompanied by boundary widening. This widening reflects increased chemical enrichment of the boundary core and the development of a more diffuse interfacial region. Notably, the temperature dependence of the boundary width in C-decorated systems is more gradual, consistent with the suppression of kite filling observed in Fig.~\ref{fig:complexions}. As a result, structural evolution in these systems occurs through a continuous broadening of the interfacial region rather than a discrete transformation.

Finally, to probe the interaction between the distinct stabilization mechanisms identified for boron and carbon, we performed targeted co-doping simulations in which boron was introduced into pre-equilibrated carbon-decorated HA grain boundaries at $x_{\mathrm{Cr}} = 25~\mathrm{at.}\%$. In these simulations, the initial configurations were first equilibrated under semi-grand canonical Monte Carlo conditions for Cr and grand canonical sampling for carbon, after which boron was introduced via grand canonical sampling while continuing active sampling of all species (Cr, C, and B) under their imposed chemical potentials. This protocol enables evaluation of boron incorporation into an already carbon-conditioned, Cr-enriched boundary state under consistent thermodynamic conditions.

Across the temperature range $300$–$1200~\mathrm{K}$, boron inserts readily into the carbon-decorated boundary, with distinct behavior observed between intermediate temperatures ($300$–$900~\mathrm{K}$) and high temperature ($1200~\mathrm{K}$). At intermediate temperatures, boron incorporation occurs without disrupting the existing carbon enrichment. Instead, boron segregation is accompanied by additional Cr accumulation at the interface. At $900~\mathrm{K}$, for example, substantial boron segregation is observed ($\Gamma_{\mathrm{B}} = 0.06~\mathrm{atoms}/\text{\AA}^2$), while $\Gamma_{\mathrm{C}}$ remains approximately constant and the Cr interfacial excess increases from $0.22$ to $0.25~\mathrm{atoms}/\text{\AA}^2$. This behavior contrasts with boron-only systems, in which boron segregation is accompanied by depletion of Cr from the boundary.

At $1200~\mathrm{K}$, boron insertion remains thermodynamically favorable but induces partial carbon depletion alongside a further increase in $\Gamma_\mathrm{Cr}$, indicating temperature-dependent redistribution among interstitial and substitutional species. Despite this redistribution, boron consistently stabilizes the co-decorated configuration, lowering $\tilde{\Omega}_{\mathrm{ex}}$ at all temperatures considered. The largest reduction occurs at $1200~\mathrm{K}$, where $\tilde{\Omega}_{\mathrm{ex}}$ decreases from $-2.29$ to $-2.74~\mathrm{J/m}^2$. These results indicate that boron remains thermodynamically driven to segregate even in carbon-conditioned, Cr-rich grain boundary environments.

\subsection*{Chemical Segregation and Short-Range Order}

\begin{figure}[H]
    \centering
    \includegraphics[width=\linewidth]{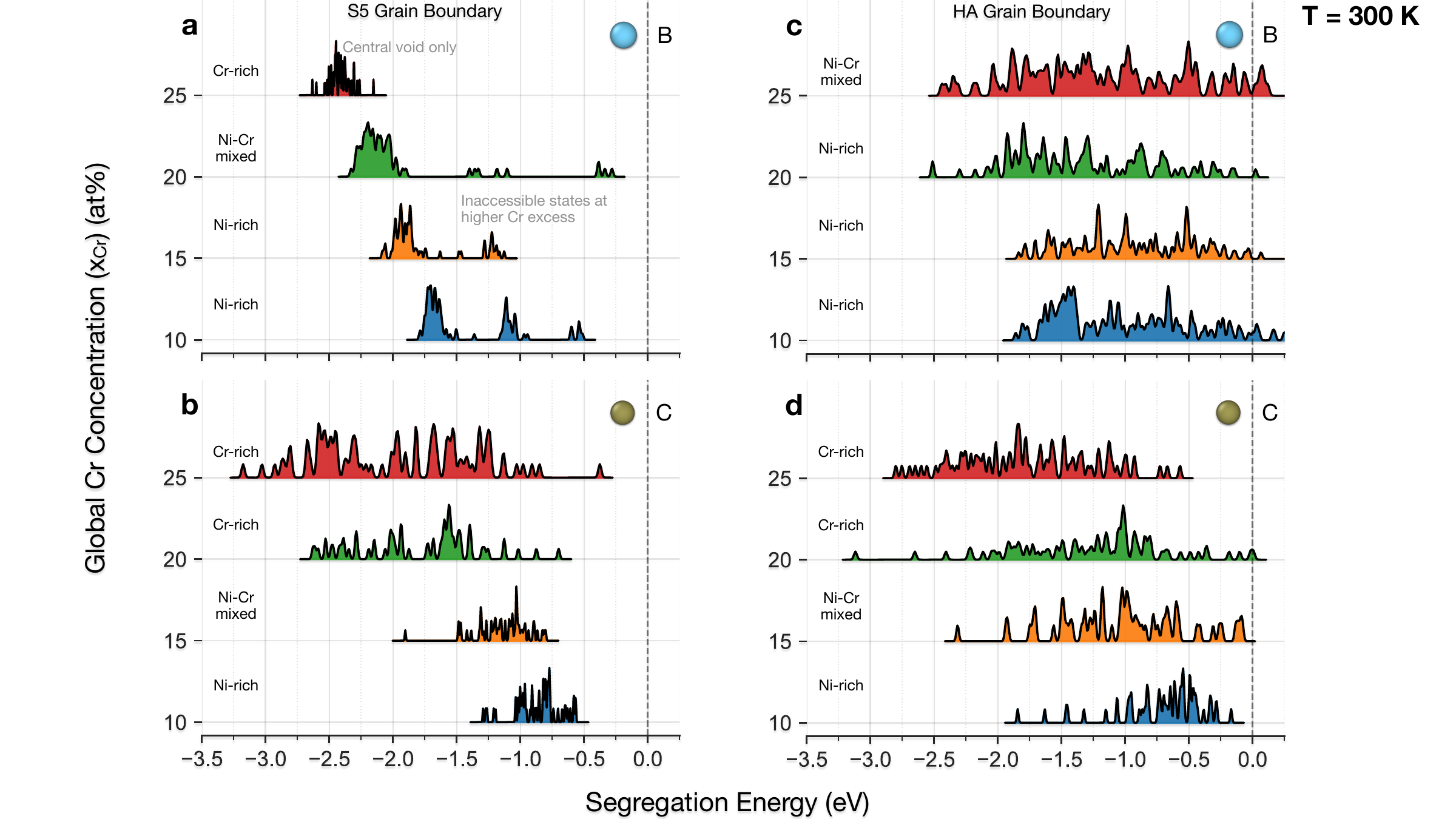}
    \caption{\textbf{Site-resolved interstitial segregation at grain boundaries.} Segregation spectra for (\textbf{a}) boron and (\textbf{b}) carbon at the $\Sigma5\,[001]\,(210)$ symmetric tilt (S5) grain boundary, and for (\textbf{c}) boron and (\textbf{d}) carbon at the high-angle (HA) grain boundary. The spectra were obtained by evaluating single-point interstitial insertions into all inequivalent sites of the equilibrated 25~at.\%~Cr grain boundary structures at 300~K. Descriptions of the grain boundary chemical state are included to connect the segregation spectra to the corresponding Cr interfacial excess ($\Gamma_{\mathrm{Cr}}$), distinguishing Ni-rich, Ni--Cr mixed, and Cr-rich boundary environments.}
    \label{fig:eseg_spectra}
\end{figure}

To further resolve how interstitial species interact with the heterogeneous energetic landscape of the grain boundary, we examine site-resolved segregation spectra for boron and carbon at the S5 and high-angle (HA) boundaries at 300~K (Fig.~\ref{fig:eseg_spectra}). For all cases, the presence of favorable segregation energies is consistent with the negative grand potential excess values observed for decorated grain boundaries.

\begin{figure}[H]
    \centering
    \includegraphics[width=\linewidth]{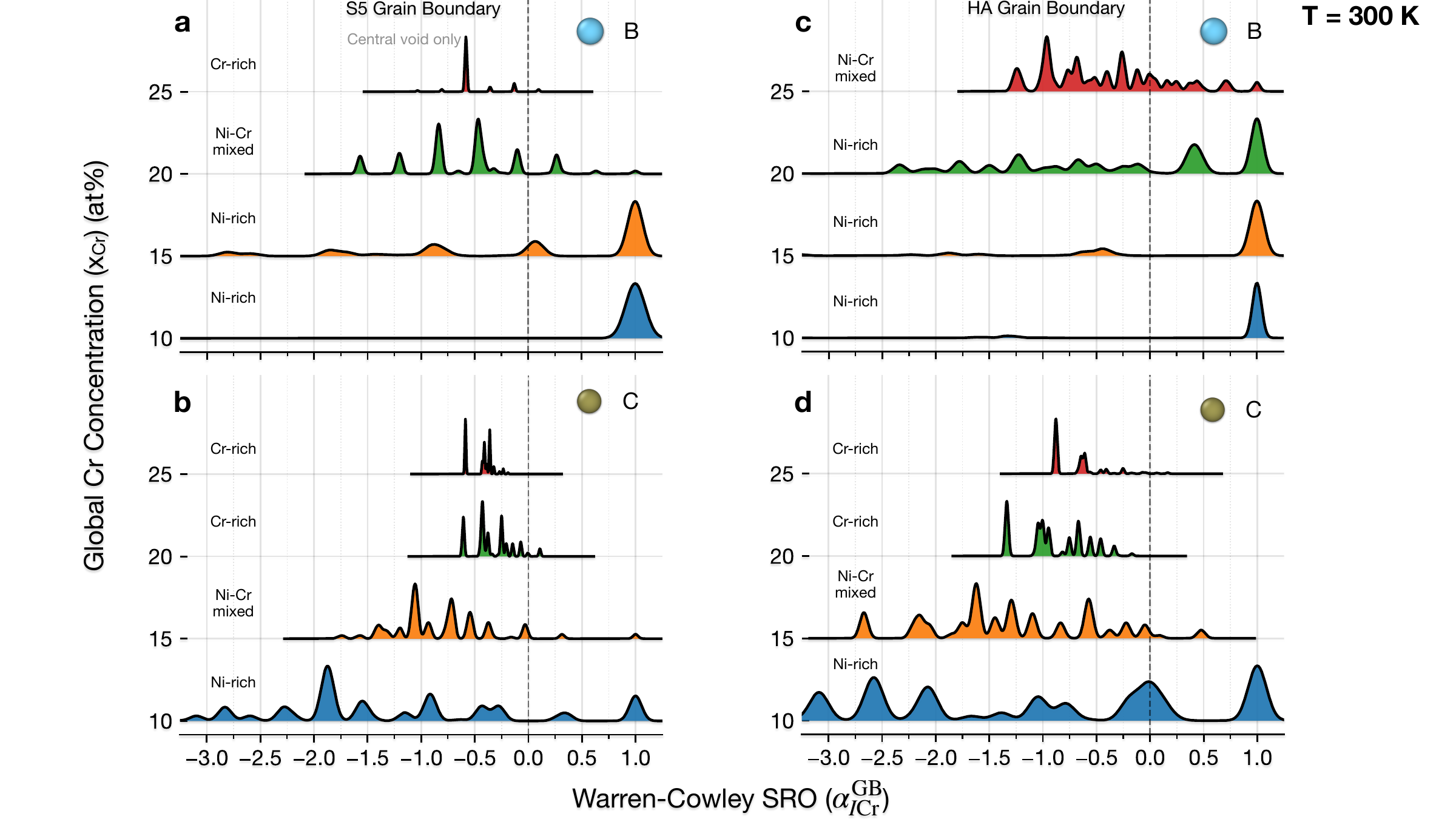}
    \caption{\textbf{Site-resolved chemical short-range order at grain boundaries.} (\textbf{a}) Boron--Cr and (\textbf{b}) carbon--Cr short-range order (SRO) spectra in the $\Sigma5\,[001]\,(210)$ (S5) grain boundary. (\textbf{c}) Boron--Cr and (\textbf{d}) carbon--Cr SRO spectra in the high-angle (HA) grain boundary. The short-range order spectra are constructed by sampling all boron or carbon atoms residing at the grain boundary in the equilibrated 25~at.\%~Cr structures. All spectra are shown for $T = 300~\mathrm{K}$ and are referenced to the local grain boundary composition.}\label{fig:sro_spectra}
\end{figure}

At the S5 boundary, boron and carbon exhibit fundamentally different spectral responses to increasing global Cr concentration. For boron, increasing $x_{\mathrm{Cr}}$ progressively eliminates energetically favorable insertion sites, as reflected by the disappearance of low-energy segregation peaks upon entering the Cr-rich regime. Only a single low-energy population associated with the central void site remains, becoming increasingly negative in energy and ultimately restricting favorable segregation to the central void shown in Fig.~\ref{fig:state_variables}a. This collapse of accessible sites provides a direct mechanistic explanation for the reduction in $\Gamma_{\mathrm{B}}$ with increasing Cr decoration. Complementary Warren--Cowley short-range order (SRO) analysis (Fig.~\ref{fig:sro_spectra}) shows that this site exhaustion is accompanied by increasingly selective B--Cr ordering at high $x_{\mathrm{Cr}}$, indicating that boron adapts to the Cr-rich boundary by localizing into the remaining Cr-tolerant motif rather than by sampling a broader population of Cr-rich environments.

Carbon exhibits the opposite response at the S5 boundary. As the boundary enters the Cr-rich regime, the segregation spectra broaden substantially, spanning approximately $-0.5$ to $-3.0$~eV. This broadening reflects the emergence of a more diverse potential energy surface driven by Cr-induced reorganization of the boundary motif and void network, as illustrated in Fig.~\ref{fig:state_variables}b. The resulting expansion of accessible insertion environments explains the monotonic increase in $\Gamma_{\mathrm{C}}$ with increasing Cr content, in direct contrast to boron behavior. Consistent with this interpretation, the SRO spectra show that while the number of C--Cr contacts increases with $x_{\mathrm{Cr}}$, the distribution of local chemical environments becomes progressively more homogeneous, leading to a narrowing of the SRO spectrum and a reduction in the fraction of strongly ordered motifs at high Cr concentrations.

At the HA boundary, broad segregation spectra are observed for both interstitial species across all Cr concentrations, consistent with the intrinsically heterogeneous nature of the disordered interface. However, the chemical response to Cr decoration remains species-dependent. Carbon exhibits a clear leftward shift of the spectra toward more negative energies with increasing $x_{\mathrm{Cr}}$, accompanied by a high density of states below $-2.0$~eV at Cr-rich conditions and the absence of positive segregation energies. In contrast, the boron spectra remain comparatively stable with increasing Cr, retaining a population of positive segregation energies even in the Cr-rich regime. SRO analysis confirms that this difference arises from carbon broadly sampling an expanding population of Cr-rich motifs, whereas boron remains constrained to a limited set of energetically favorable sites that tolerate, but do not actively recruit, Cr.

Together, these results demonstrate that segregation spectra encode the accessibility of insertion sites, while SRO spectra resolve how chemical ordering evolves within those sites. The combined analysis reveals distinct stabilization pathways for boron and carbon that depend on both interstitial identity and grain boundary character. Quantitative spectral descriptors supporting these trends are provided in Figs.~S7~and~S8 of the Supplemental Materials.

\section*{Discussion}

\begin{figure}[H]
    \centering
    \includegraphics[width=\linewidth]{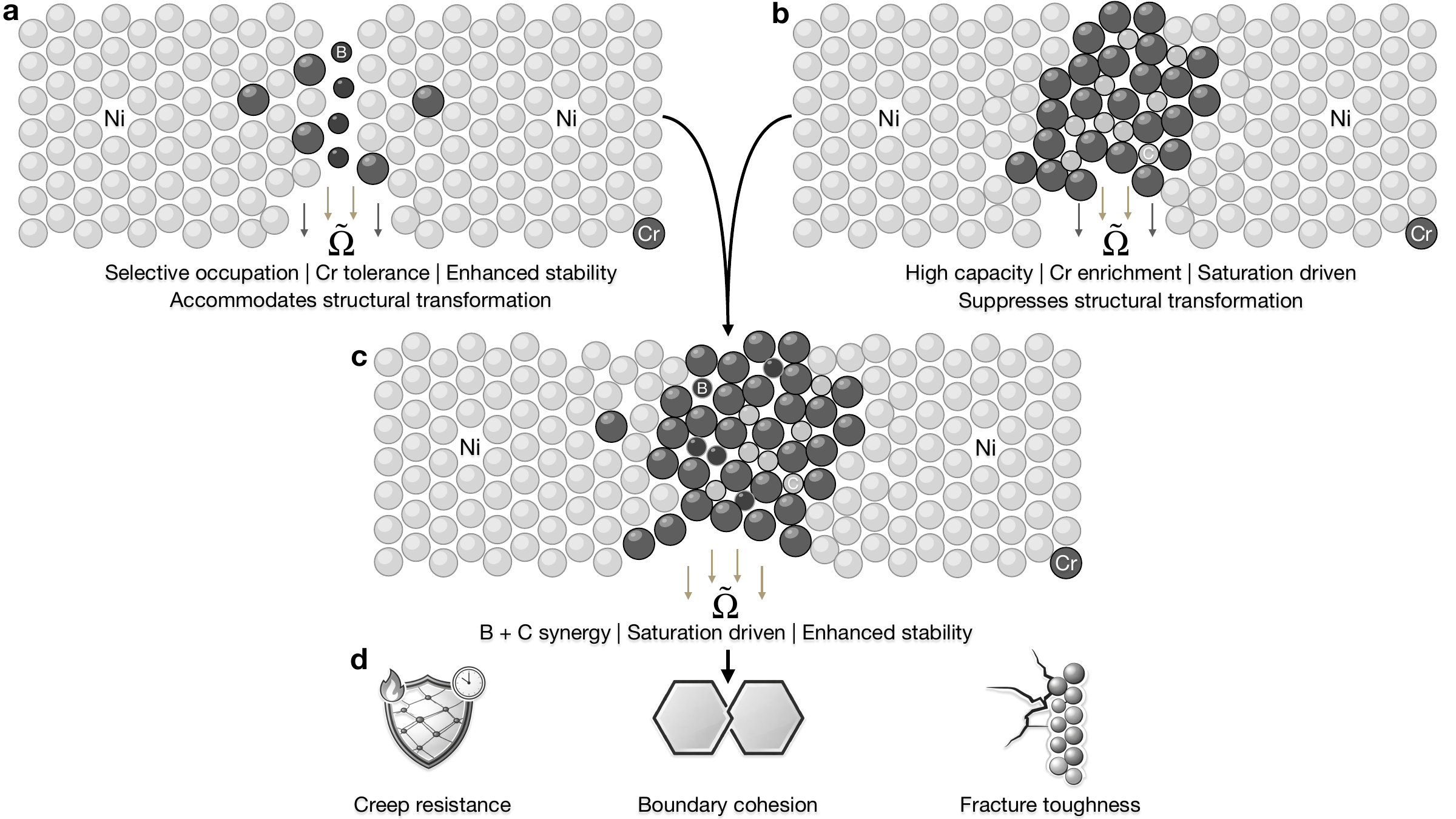}
    \caption{\textbf{Schematic illustration of interstitial stabilization mechanisms at grain boundaries.}
(a) Boron stabilizes boundaries through selective site occupation and tolerance of Cr co-segregation, promoting gradual structural evolution of the boundary. 
(b) Carbon promotes high-capacity segregation and Cr enrichment through saturation-driven filling of interfacial sites, stabilizing grain boundary structures and suppressing temperature-driven structural transitions. 
(c) Targeted co-doping simulations indicate that carbon conditions the boundary chemically while boron remains thermodynamically driven to segregate, further lowering the interfacial grand potential without displacing existing segregation. 
(d) Implications for macroscopic properties, including creep resistance, boundary cohesion, and fracture toughness.}\label{fig:synthesis}
\end{figure}

The results presented here reveal a coupled chemo-structural response of grain boundaries to interstitial chemistry and temperature. As demonstrated in Fig.~\ref{fig:complexions}, the S5 grain boundary exhibits a temperature-dependent structural transition from open to filled kite configurations, while the high-angle boundary undergoes a more diffuse structural evolution characterized by progressive broadening of the interfacial region. These results show that interstitial species not only modify segregation behavior but also reshape the accessible structural states of the boundary, with boron promoting gradual structural transitions and carbon delaying or suppressing transformation over an extended temperature range.

Within this structural context, the thermodynamic analysis provides a quantitative framework for comparing the stability of these chemo-structural states under fixed chemical potentials of light interstitials. Boron and carbon stabilize grain boundaries through distinct but complementary mechanisms. Carbon conditions the boundary by promoting Cr-rich interfacial environments through saturation-driven co-segregation, particularly in structurally disordered boundaries where carbon accesses a broad population of deep segregation sites. In contrast, boron stabilizes the boundary through selective occupation of energetically favorable interfacial sites while tolerating Cr co-occupancy, leading to consistently lower interfacial grand potential excess without requiring substantial Cr enrichment.

Targeted co-doping simulations further show that boron remains thermodynamically driven to segregate even in carbon-conditioned, Cr-rich grain boundary environments. In these simulations, carbon establishes a chemically enriched boundary state that remains accessible to subsequent boron incorporation, with boron continuing to lower the interfacial grand potential without displacing carbon. These observations indicate that boron and carbon are not inherently competitive at the interface, but can coexist within chemically enriched grain boundary states.

These stabilization pathways do not imply the direct formation or suppression of specific interfacial compounds, nor do they predict precipitation kinetics or time-dependent transformation pathways. Instead, they quantify how interstitial identity and substitutional chemistry reshape the energetic landscape and modulate the accessible structural states of the boundary, including temperature-dependent transitions between distinct grain boundary structures. This framework provides a basis for understanding why co-doping strategies can outperform single-interstitial approaches across structural alloy systems and suggests quantitative design principles for grain boundary engineering, summarized schematically in Fig.~\ref{fig:synthesis}.

Figs.~\ref{fig:state_variables} and \ref{fig:eseg_spectra} show that interfacial stability, chemical enrichment, and structural width cannot be described by a single monotonic or universal relationship with interstitial chemistry, as further clarified by local chemical ordering trends discussed below. Instead, boron and carbon operate through qualitatively distinct thermodynamic modes that reflect fundamentally different interactions with the grain boundary energy landscape and, correspondingly, give rise to different accessible structural states of the boundary. Critically, these responses do not arise from isolated interstitial effects but from the coupled thermodynamics of interstitial–substitutional co-segregation. The presence of Cr as a substitutional alloying element reshapes the segregation response of both boron and carbon, producing concentration-dependent competition and synergy that cannot be inferred from interstitial behavior in pure Ni or Fe matrices.

Boron acts as a selective interfacial stabilizer, producing a consistent reduction in the interfacial grand potential excess through preferential stabilization of energetically favorable configurations rather than extensive chemical enrichment. This behavior is observed for both S5 and HA grain boundaries despite low interfacial excess values. In S5 boundaries, increasing Cr content progressively restricts boron segregation to fewer thermodynamically accessible environments, reflected by the loss of shallow segregation populations and the survival of only the deepest insertion sites in the segregation spectra (Fig.~\ref{fig:eseg_spectra}). Consistent with this site exhaustion, SRO spectra (Fig.~\ref{fig:sro_spectra}) show that boron samples an increasingly narrow distribution of local chemical environments at high $x_{\mathrm{Cr}}$, reflecting localization into a small set of Cr-rich motifs. In contrast, HA boundaries retain a broad distribution of accessible sites, with boron sampling diverse chemical environments while deriving stabilization from energetically favorable Cr-rich configurations.

This behavior reflects a thermodynamic partitioning effect arising from interstitial substitutional interactions. At moderate Cr concentrations, boron segregation depletes Cr from the boundary, as evidenced by strongly negative $\Gamma_{\mathrm{Cr}}$ values across most conditions (Fig.~\ref{fig:state_variables}e). At elevated Cr concentrations (25~at.\%), the imposed chemical potential prevents complete Cr expulsion, forcing coexistence of boron and Cr at the interface. SRO analysis shows that this crossover corresponds to a transition from boron occupying predominantly Ni-rich motifs at low $x_{\mathrm{Cr}}$ to selective localization within Cr-adjacent sites at high $x_{\mathrm{Cr}}$, explaining the increasingly negative B–Cr ordering despite declining boron populations. Boron remains favorably segregated because the thermodynamic penalty associated with B–Cr interactions increases more rapidly in the bulk alloy than at the grain boundary~\cite{dolezalSpectralAnalysisLight2026a}, indicating that boron tolerates Cr co-occupancy more readily at the interface than in the crystalline bulk.

In contrast, carbon stabilizes grain boundaries through a saturation-driven mechanism characterized by strong Cr co-enrichment, high interfacial excess, and progressive broadening of the interfacial region. Carbon segregation is accompanied by substantial Cr accumulation at the boundary, reaching interfacial excess values of approximately 0.225~atoms/\AA$^2$ compared to 0.05~atoms/\AA$^2$ for boron-decorated boundaries at 25~at.\% Cr (Fig.~\ref{fig:state_variables}e). This Cr co-enrichment is not incidental but represents the defining feature of carbon's stabilization mechanism, a trend reflected in SRO spectra by broad ordering distributions at low Cr content that progressively narrow as Cr-rich motifs become ubiquitous rather than exceptional. Carbon thus acts as a chemical driver that actively recruits Cr to the interface, producing Cr-rich boundary environments.

The corresponding segregation spectra reveal access to a wide distribution of interfacial environments that broaden systematically with increasing Cr content (Fig.~\ref{fig:eseg_spectra}), consistent with a capacity-dominated stabilization mode in which boundary reinforcement scales with chemical occupation rather than energetic selectivity. This saturation-driven behavior is further reflected in the SRO spectra, which show that carbon initially samples a wide range of chemically distinct environments at low Cr content, followed by progressive homogenization as Cr-rich motifs become prevalent. In conjunction with the structural analysis in Fig.~\ref{fig:complexions}, this behavior is associated with suppression of discrete structural transformations and the persistence of comparatively less-evolved, more spatially confined boundary configurations over a broad temperature range. 

Such Cr-enriched interfacial chemistries are widely observed experimentally as M$_{23}$C$_6$ carbides, the dominant grain-boundary carbide phase in Ni-based superalloys during intermediate-temperature exposure~\cite{simsSuperalloysIIHighTemperature1987,kontisEffectBoronGrain2016,duPrecipitationEvolutionGrain2017a}. In this regime, stability arises from broad compositional reorganization and chemical saturation rather than selective minimization of the interfacial energy landscape. Once carbon-conditioned, Cr-rich boundaries are established, strong segregation of boron to Cr-rich M$_{23}$C$_6$ carbide-decorated grain boundaries has been reported in Ni-based superalloys~\cite{kontisEffectBoronGrain2016,tytkoMicrostructuralEvolutionNibased2012a}. This behavior suggests that boron preferentially stabilizes boundaries that have undergone carbon-mediated chemical conditioning. This interpretation is supported by the targeted co-doping simulations at $x_\mathrm{Cr} = 25~\mathrm{at.}\%$ and $300$--$1200~\mathrm{K}$. Carbon enrichment remained stable while boron inserted readily and further increased $\Gamma_\mathrm{Cr}$. This yielded a more favorable interfacial state, as $\tilde{\Omega}_\mathrm{ex}$ became more negative at all temperatures, confirming that boron stabilizes an existing carbon-conditioned, Cr-rich interface rather than displacing it.

These two modes, selective versus saturation-driven stabilization, explain why increased chemical enrichment does not necessarily translate into improved thermodynamic stability. This distinction arises directly from the interplay between interstitial chemistry and substitutional alloying: carbon's strong affinity for Cr drives saturation-driven enrichment and extensive chemical reorganization, whereas boron's tolerance for Cr enables selective site occupation that stabilizes the interface without inducing broad compositional restructuring.

The present results demonstrate that a lower interfacial grand potential excess can be achieved through ordering-dominated mechanisms that require neither high solute content nor significant grain boundary widening. As resolved by SRO analysis, this ordering-dominated stabilization reflects selective localization of interstitials into energetically favorable chemical motifs rather than broad chemical homogenization. This distinction persists across boundary character. Even in disordered high-angle boundaries that accommodate large interstitial excesses and broad energetic distributions, boron consistently yields a more favorable interfacial grand potential than carbon under identical thermodynamic conditions. Together with the co-doping results, this behavior shows that thermodynamic stabilization is governed not by the magnitude of chemical enrichment alone, but by how interstitial--substitutional interactions reshape site accessibility, interfacial order, and the range of accessible structural states.

The extent to which selective and saturation-driven stabilization modes can be expressed is further modulated by grain boundary character, with ordered boundaries amplifying energetic selectivity and favoring more discrete structural transitions, while disordered boundaries favor capacity-driven enrichment and more continuous structural evolution.

Experimental observations in Ni-based superalloys indicate that boron operates in distinct regimes depending on concentration: as an elemental segregant at low activities and as a boride-forming species at elevated activities~\cite{kontisEffectBoronGrain2016,duPrecipitationEvolutionGrain2017a,garosshenEffectsZrStructure1987,tytkoMicrostructuralEvolutionNibased2012a, kontisInfluenceCompositionPrecipitation2019}. At low boron concentrations ($\sim$0.01~at.\% and below), boron segregates to $\gamma$/M$_{23}$C$6$ and $\gamma'$/M${23}$C$_6$ interfaces without forming boride phases, enhancing cohesion within carbon-conditioned, Cr-rich boundary regions. At elevated concentrations ($\sim$0.03~at.\% and above), M$_5$B$3$ boride formation occurs, disrupting continuous M${23}$C$_6$ morphologies and producing discrete carbide and boride pockets along grain boundaries. While these regimes differ fundamentally in their interfacial chemistry and microstructural manifestation, both are consistently associated with improved high-temperature creep resistance relative to boron-free alloys, indicating that boron enhances grain boundary performance whether present as an elemental segregant or incorporated into boride phases.

The selective-saturation framework developed here provides a thermodynamic basis for interpreting these experimentally observed transitions. Boron's comparatively weaker interaction with Cr enables selective stabilization at low enrichment levels without driving saturation-driven chemical reorganization, while carbon's broad segregation spectra and strong Cr affinity promote saturation-driven enrichment that establishes the chemical template for carbide formation. The concentration-dependent behavior observed experimentally highlights that boron--carbon cooperation is tunable rather than fixed, with the optimal balance depending on alloy chemistry, service temperature, and targeted mechanical response, as these factors collectively determine both the chemical state and the accessible structural configurations of the grain boundary.

The relevance of this framework extends beyond Ni-based superalloys. In steels, carbon-driven segregation establishes Mn- and Cr-enriched grain boundary phases~\cite{xuCharacterizationM23C6Carbides2016}, within which boron has been shown to influence grain boundary structure, enhance cohesion, and modify hydrogen embrittlement behavior through segregation to carbon-conditioned interfaces~\cite{zhouBoronTriggersGrain2025,dongDualRoleBoron2025}. These observations are consistent with the complementary stabilization behaviors identified here, suggesting that similar interstitial interactions may operate across a broader range of alloy systems. Direct computational evidence for boron–carbon synergy in Ni-based superalloys has been demonstrated through density functional theory studies of sulfur embrittlement resistance~\cite{chenEnhancingSulfurEmbrittlement2025b}. In that work, boron and carbon co-doping produced strengthening effects exceeding those of individual additions.

Beyond boron–carbon combinations, recent experimental studies indicate that cooperative interstitial strategies are already emerging in alloy design. In additively manufactured Inconel~718, boron and phosphorus act as complementary dopants: neither element alone significantly improves creep performance, whereas their combined addition produces marked gains in creep ductility and rupture life, accompanied by a transition from predominantly intergranular fracture to mixed fracture modes~\cite{wangImprovedCreepProperties2024a}. Long-term exposure studies further show that boron stabilizes interfacial states only after phosphorus-driven solute redistribution~\cite{wangEffectsPhosphorusBoron2024}. Similar cooperative behavior has been reported in high-entropy alloys, where boron–nitrogen co-doping produces synergistic strengthening through concurrent grain boundary segregation and precipitation hardening~\cite{sonFacileStrengtheningMethod2022a}. These results indicate a broader shift away from single-dopant paradigms toward deliberately engineered interstitial combinations.

The mechanistic framework developed here provides a basis for thermodynamic design principles for grain boundary engineering in substitutionally complex alloys. Effective stabilization cannot be achieved by treating light interstitials as independent strengthening agents operating in isolation from substitutional chemistry. Instead, stabilization arises from three coupled principles reflected in our results: carbon establishes Cr-rich boundary environments through saturation-driven enrichment and promotes structurally diffuse interfacial states while suppressing temperature-driven structural transitions; boron stabilizes these environments by selectively lowering the interfacial grand potential through localized chemical ordering while enabling gradual structural evolution; and the interplay between interstitial segregation and Cr redistribution governs site accessibility, chemical competition, and the range of accessible structural configurations of the boundary. While prior investigations of boron or carbon in pure Ni or Fe matrices provide foundational insight, they cannot capture the concentration-dependent competition and interaction that arise when multiple interstitial species interact with substitutional elements such as Cr. Together, these mechanisms provide a thermodynamic basis for understanding the coupled chemo-structural stability associated with the enhanced mechanical performance observed experimentally in boron--carbon--containing structural alloys under demanding service conditions.


\section*{Methods}

\subsection*{Semi-Grand Canonical Monte Carlo Sampling}

To capture both local chemical rearrangements and global composition changes, two types of Monte Carlo moves were employed. First, the metallic swap move is a canonical ($NVT$) exchange between two lattice sites occupied by different metallic species and is accepted according to the standard Metropolis criterion \cite{metropolisEquationStateCalculations1953},

\begin{equation}
P^\mathrm{swap}_\text{accept} =
\min\left[1,\exp\!\left(-\beta\Delta U\right)\right],
\end{equation}
where $\Delta U$ is the potential energy difference between the final and initial states. Because the total composition is conserved during a metallic swap, this move samples local chemical rearrangements without coupling to an external chemical potential reservoir. It was included to accelerate equilibration of the chemical short-range order (SRO) among substitutional species during semi-grand canonical equilibration.

To simulate global chemical partitioning between Ni and Cr, we used the semi-grand canonical ($N,V,T,\Delta \mu$) Monte Carlo (SGC-MC) formalism. In this ensemble, the number of lattice sites remains fixed while transmutation moves are attempted between metallic species Ni~$\leftrightarrow$~Cr. The acceptance probabilities for forward and reverse transmutations are given by

\begin{align}
P^{\mathrm{Ni}\to\mathrm{Cr}}_\text{accept} &= \min\!\left[1, \exp\!\left(-\beta[\Delta U - \Delta\mu_{\mathrm{Cr}\rightarrow\mathrm{Ni}}]\right)\right], \label{eq:transmute_forward}\\[10pt]
P^{\mathrm{Cr}\to\mathrm{Ni}}_\text{accept} &= \min\!\left[1, \exp\!\left(-\beta[\Delta U + \Delta\mu_{\mathrm{Cr}\rightarrow\mathrm{Ni}}]\right)\right], \label{eq:transmute_reverse}
\end{align}
where $\beta = 1/(k_\mathrm{B}T)$ and $\Delta\mu_{\mathrm{Cr}\rightarrow\mathrm{Ni}} = \mu_\mathrm{Cr} - \mu_\mathrm{Ni}$ is the imposed chemical potential difference between the two species. By varying $\Delta\mu_{\mathrm{Cr}\rightarrow\mathrm{Ni}}$, one can reproduce the equilibrium Ni--Cr composition at each temperature for a chosen target solute concentration $x_\mathrm{Cr}$. Bulk SGC-MC simulations were performed to calibrate $\Delta\mu_{\mathrm{Cr}\rightarrow\mathrm{Ni}}(x_\mathrm{Cr},T)$ values corresponding to $x_\mathrm{Cr} =$~10--25~at.\% (in 5~at.\% increments) at 300, 600, 900, and 1200~K. During these calibrations, $\Delta\mu_{\mathrm{Cr}\rightarrow\mathrm{Ni}}$ was iteratively adjusted until $\langle x_\mathrm{Cr} \rangle$ equilibrated within 0.2~at.\% of the target value. The resulting $\Delta\mu_{\mathrm{Cr}\rightarrow\mathrm{Ni}}(x_\mathrm{Cr},T)$ values were then used as fixed input parameters for subsequent binary alloy grain boundary simulations. The selection of Cr as the co-segregant is motivated by experimental evidence showing that both boron and carbon co-segregate with Cr~\cite{cadelAtomProbeTomography2002a, tytkoMicrostructuralEvolutionNibased2012a, tekogluSuperiorHightemperatureMechanical2024d, kontisEffectBoronGrain2016, kontisInfluenceCompositionPrecipitation2019, duPrecipitationEvolutionGrain2017a}.

\subsection*{Hybrid Semi-Grand/Grand Canonical Monte Carlo Sampling}

Interstitial solutes (boron and carbon) were modeled using a hybrid SGC/grand canonical (GC) scheme in which metallic substitutions were treated semi-grand canonically, while interstitial insertion and deletion moves were attempted grand canonically. For trial insertion and deletion of an interstitial atom $I$, the acceptance probabilities are given by  
\begin{align}
P_I^\mathrm{insert} &= 
\min\!\left[1,\, \frac{V}{\Lambda^3 (N_I + 1)} 
\exp\!\left(-\beta[\Delta U - \mu^\mathrm{ext}_I]\right)\right], 
\label{eq:insert}\\[10pt]
P_I^\mathrm{delete} &= 
\min\!\left[1,\, \frac{\Lambda^3 N_I}{V} 
\exp\!\left(-\beta[\Delta U + \mu^\mathrm{ext}_I]\right)\right],
\label{eq:delete}
\end{align}
where $\Delta U$ is the potential energy difference between the final and initial states, 
$V$ is the simulation cell volume, $\Lambda$ is the thermal de~Broglie wavelength, 
$\mu^\mathrm{ext}_I$ is the imposed chemical potential of species $I$, and $N_I$ is the number of interstitials of type $I$ currently present. Insertion and deletion moves were attempted with equal probability to preserve detailed balance.

The mean site occupancy of an interstitial follows the Fermi-Dirac form derived in the Supplemental Materials (Section~4),
\begin{equation}\label{eq:theta}
\theta_I = \frac{1}{1 + \exp\!\left(\frac{E_0 - \mu^\mathrm{ext}_I}{k_\mathrm{B} T}\right)},
\end{equation}
where $E_0$ is the insertion free energy of the interstitial species (boron or carbon) in a reference octahedral site of bulk Ni at the same temperature and pressure. 
This expression treats interstitials as dilute, non-interacting point defects in a rigid, substitutionally equilibrated host lattice, conditions appropriate for the low concentrations considered here. 

Inverting Eq.~\ref{eq:theta} gives the external chemical potential required to impose a desired bulk occupancy,
\begin{equation}
\mu^\mathrm{ext}_I = E_0 + k_\mathrm{B}T
\ln\!\left(\frac{\theta_I}{1 - \theta_I}\right)
\simeq E_0 + k_\mathrm{B}T\ln\theta_I ,
\end{equation}
where the dilute-limit approximation is valid for $\theta_I \ll 1$. 
Here, target bulk occupancies of $\theta_\mathrm{B} = 10^{-4}$ (0.01~at.\%) and $\theta_\mathrm{C} = 10^{-3}$ (0.1~at.\%) were adopted to represent typical experimental dopant levels in of boron and carbon Ni-based alloys~\cite{kontisEffectBoronGrain2016, duPrecipitationEvolutionGrain2017a,garosshenEffectsZrStructure1987, kontisInfluenceCompositionPrecipitation2019}. The final values used were $\mu^\mathrm{ext}_\mathrm{B}=-6.70$~eV and $\mu^\mathrm{ext}_\mathrm{C}= -7.69$~eV for all ($x_\mathrm{Cr}$, $T$) conditions.
These values define composition- and temperature-independent chemical potentials $\mu_\mathrm{B/C}^{\mathrm{ext}}$ such that all variations in segregation behavior arise solely from local changes in Cr content or interfacial structure, rather than from shifts in the bulk reservoir. In total, four types of MC trial moves were employed: (1) substitutional transmutation, (2) interstitial insertion, (3) interstitial deletion, and (4) metallic atom swap.

\subsection*{Simulation Protocol}

(Step 1) SGC-MC simulations for $\Delta\mu_{\mathrm{Cr}\rightarrow\mathrm{Ni}}$ convergence were performed using a $10\times10\times10$ face-centered cubic (FCC, Fm$\bar{3}$m) supercell of Ni--Cr containing 4,000 atoms. Details of the hierarchical convergence workflow and error scaling analysis are provided in Section~3 of the Supplemental Materials. (Step 2) Hybrid SGC/GC-MC simulations were carried out on two grain boundary structures: (i) the $\Sigma5\,[001]\,(210)$ symmetric-tilt boundary (S5), a well-studied archetype for FCC metals~\cite{tschoppSymmetricAsymmetricTilt2015,zhengGrainBoundaryProperties2020}; 
and (ii) a custom high-angle boundary (HA) designed to represent the diverse spectrum of boundaries typically present in polycrystalline materials~\cite{rohrerComparingCalculatedMeasured2010,schuhUniversalFeaturesGrain2005a}. The chosen HA grain boundary structure corresponds to a non-CSL boundary with a misorientation angle of 106.5$^\circ$. Each grain boundary simulation cell was initialized and randomly decorated according to the target solute atomic fraction $x_\mathrm{Cr}$. All post-MC configurations were subsequently annealed at the target temperature for 50~ps using $NVT$ MD with a 1~fs timestep. This annealing step serves to: (1) relax residual stress fields introduced by discrete MC insertion/deletion events, (2) allow local structural equilibration of the grain boundary, and (3) detect potential thermally-activated complexion transitions that may occur on accessible MD timescales but were not sampled during the MC phase due to kinetic barriers or collective atomic rearrangements.

Schematics of the undoped pure Ni grain boundary structures are shown in Fig.~S1 of the supplemental materials. The calculated grain boundary energies for the undoped pure Ni reference structures were 1.20~J\,m$^{-2}$ for S5 and 1.74~J\,m$^{-2}$ for HA. This pairing was selected because low-$\Sigma$ symmetric tilts, such as $\Sigma5$, sample only a narrow subset of the polycrystalline boundary landscape and therefore do not capture the full structural or chemical heterogeneity of real materials~\cite{schuhUniversalFeaturesGrain2005a,wagihViewpointCanSymmetric2023}. Full details of the HA structure generation procedure are provided in Section~2 of the Supplemental Materials.

The potential energy difference for each trial configuration was evaluated using the Atomic Simulation Environment (ASE)~\cite{hjorthlarsenAtomicSimulationEnvironment2017} with version~5.0.0 of the universal core neural network Preferred Potential (PFP)~\cite{takamotoUniversalNeuralNetwork2022} including the D3 dispersion correction implemented through Matlantis~\cite{Matlantis}. Structural relaxations were performed using the fast inertial relaxation engine (FIRE) algorithm with a force convergence criterion of 0.01~eV\,\AA$^{-1}$. All thermodynamic quantities reported in this work, including the transmutation chemical potentials ($\Delta\mu_{\mathrm{Cr}\rightarrow\mathrm{Ni}}$) and interstitial reservoir potentials ($\mu_\mathrm{B/C}^{\mathrm{ext}}$), were derived consistently using total energies obtained from the same PFP model. The accuracy and transferability of the PFP framework have been extensively validated by its developers and in our previous works~\cite{dolezalSegregationOrderingLight2025, dolezalAtomisticMechanismsOxidation2025}. For completeness, a summary of DFT-to-PFP validation results are provided in the Supplemental Materials (Section~8).

\subsection*{Interfacial state variables}

Grain boundaries were modeled as low-dimensional thermodynamic systems whose equilibrium states can be parameterized by a minimal set of state variables capturing both chemical and structural deviations from the surrounding bulk. These variables form the basis of an interfacial state, wherein distinct boundary states are mapped as functions of temperature and global alloy composition. The chosen state variables include the grand potential excess energy per grain boundary area ($\tilde{\Omega}_\mathrm{ex}$), Gibbsian interfacial excess ($\Gamma$), and the post-equilibration boundary width ($w$). 

The relative stability of each equilibrated configuration was quantified using the grand potential excess energy per unit grain boundary area, referenced to the corresponding pure Ni grain boundary. The grand potential excess is defined as  
\begin{equation}\label{eq:omega_gb}
\tilde{\Omega}_\mathrm{ex}(x_\mathrm{Cr}, T) =
\frac{E_\mathrm{GB}(x_\mathrm{Cr}, T) - E_\mathrm{GB}^{\mathrm{Ni,ref}}
- N_\mathrm{Cr}\,\Delta\mu_{\mathrm{Cr}\rightarrow\mathrm{Ni}}(x_\mathrm{Cr}, T)
- N_I \mu^\mathrm{ext}_I}{2A_\mathrm{GB}},
\end{equation}
where $E_\mathrm{GB}(x_\mathrm{Cr},T)$ is the total energy of the relaxed bicrystal supercell at global Cr fraction $x_\mathrm{Cr}$ and temperature $T$, and $E_\mathrm{GB}^{\mathrm{Ni,ref}}$ is the total energy of the corresponding pure Ni grain boundary supercell. The reference system shares the same bicrystal geometry, crystallographic character, and simulation protocol (including MC equilibration and MD annealing) as the interstitial-doped alloyed configurations, ensuring that $\tilde{\Omega}_\mathrm{ex}$ reflects changes in interfacial thermodynamic stability arising from chemical substitution and interstitial incorporation under consistent structural and thermodynamic conditions. 

Importantly, the reference grain boundary structure is temperature-dependent, reflecting the equilibrium configuration of the pure Ni boundary at each temperature (open kite configurations at 300~K and filled kite configurations at elevated temperatures following structural relaxation). As a result, $\tilde{\Omega}_\mathrm{ex}$ provides a measure of relative stability with respect to the appropriate structural state of the boundary under each thermodynamic condition. Negative values of $\tilde{\Omega}_\mathrm{ex}$ indicate enhanced stability relative to the reference grain boundary, while positive values indicate reduced stability, and do not correspond to an absolute driving force for grain boundary formation.

$N_\mathrm{Cr}$ denotes the number of Cr atoms in the simulation cell, $N_I$ is the number of interstitial atoms of type $I \in \{\mathrm{B},\mathrm{C}\}$, and $A_\mathrm{GB}$ is the total grain boundary area. All energies are time-averaged over the final 25~ps of the equilibrated canonical ($NVT$) molecular dynamics trajectory. The term $\Delta\mu_{\mathrm{Cr}\rightarrow\mathrm{Ni}}$ represents the chemical potential difference between Cr and nickel in the bulk alloy reservoir at the specified composition and temperature, while $\mu^\mathrm{ext}_I$ is the externally imposed interstitial chemical potential used consistently in the hybrid SGC/GC-MC simulations.

While Monte Carlo sampling primarily explores chemical configurations, the subsequent molecular dynamics annealing step enables local structural relaxation of the grain boundary and captures thermally activated structural rearrangements, including temperature-driven transitions that are kinetically accessible on the MD timescale. As such, the computed grand potential reflects a coupled chemo-structural equilibration within the set of grain boundary states sampled during the simulations. However, the present framework does not explicitly or exhaustively sample all possible grain boundary phases, and the resulting configurations should therefore be interpreted as locally equilibrated states within a restricted region of grain boundary structural phase space.

The Gibbsian interfacial excess of species~$i$ was evaluated by comparing the atomic population within the grain boundary region to that of the bulk reference:
\begin{equation}\label{eq:excess}
\Gamma_i = \frac{N_i^\mathrm{GB}}{A_\mathrm{GB}} 
- c_i^{\mathrm{bulk}} \frac{N^\mathrm{GB}}{A_\mathrm{GB}},
\end{equation}
where $N_i^\mathrm{GB}$ and $N^\mathrm{GB}$ are the numbers of atoms of species~$i$ and the total atoms within the grain boundary region, respectively, $c_i^{\mathrm{bulk}}$ is the bulk atomic fraction of species~$i$, and $A_\mathrm{GB}$ is the interfacial area. Positive values of $\Gamma_i$ correspond to interfacial segregation, while negative values indicate depletion. 

The grain boundary width was obtained from the spatial distribution of atoms identified as ``other'' (non-fcc) by common-neighbor analysis. The fraction of such atoms was binned along the grain boundary normal direction ($\hat{y}$) and fit to a finite-width step-wave profile constructed from a difference of two error functions,
\begin{equation}
\phi(y) = \frac{A}{2}
\left[
\mathrm{erf}\!\left( \frac{y - \left(y_0 - \frac{w}{2}\right)}{\sigma_{\mathrm{int}}\sqrt{2}} \right)
-
\mathrm{erf}\!\left( \frac{y - \left(y_0 + \frac{w}{2}\right)}{\sigma_{\mathrm{int}}\sqrt{2}} \right)
\right],
\end{equation}
where $y_0$ denotes the grain boundary center, $A$ represents the plateau amplitude of the disordered region, $w$ is the grain boundary width defined as the separation between the midpoints of the two interfacial transitions, and $\sigma_{\mathrm{int}}$ characterizes the interfacial broadening at the boundary edges. This formulation captures a structurally disordered interfacial region of finite thickness bounded by diffuse interfaces. Together, $\{\tilde{\Omega}_{\mathrm{ex}}, \Gamma_i, w\}$ define the thermodynamic coordinates of each equilibrated configuration.

\subsection*{Segregation Energy Spectra}
The variability of segregation behavior across inequivalent grain boundary sites was characterized through spectral analysis~\cite{wagihSpectrumGrainBoundary2019,wagihLearningGrainBoundarySegregation2022}. In this approach, the distribution of segregation energies across all available interstitial sites defines a segregation spectrum, which captures the statistical heterogeneity of local atomic environments within a boundary. Such spectra provide a physically rigorous basis for comparing grain boundaries of different character and for quantifying how temperature and composition alter the energy landscape accessible to solute atoms. 

To construct the segregation spectra, all equilibrated post-MC/MD configurations at $T=300$~K were first stripped of interstitials, returning each boundary to its undoped state. A single interstitial atom (boron or carbon) was then inserted sequentially into every previously occupied interstitial position, and the resulting segregation energies, $E_\mathrm{seg}$, were evaluated. The ensemble of $E_\mathrm{seg}$ values obtained in this manner defines the full segregation spectrum for that grain boundary. The segregation energy is defined in Eq.~\ref{eq:eseg},

\begin{equation}\label{eq:eseg}
    E^{(I)}_\mathrm{seg} = E^{I@\mathrm{GB}}_\mathrm{GB} - \langle E^{I@\mathrm{bulk}}_\mathrm{GB}\rangle
\end{equation}
where $E^{I@\mathrm{GB}}_\mathrm{GB}$ is the total energy of the system with an interstitial occupying a grain boundary void site, and $\langle E^{I@\mathrm{bulk}}_\mathrm{GB} \rangle$ is the total energy of the system with the interstitial placed in an octahedral site within the bulk region of the same simulation cell. The angular brackets denote averaging over multiple bulk sites, which accounts for local variations in Cr content within the first nearest-neighbor coordination shell of each octahedral cage. Negative $E^{(I)}_\mathrm{seg}$ values indicate energetically favorable segregation to the grain boundary, while positive values indicate preference for bulk occupancy.

\subsection*{Short-Range Order Spectra}

Local chemical short-range order (SRO) was quantified using the Warren--Cowley (WC) SRO parameter
\cite{cowleyApproximateTheoryOrder1950, cowleyShortLongRangeOrder1960}. For an interstitial--solute pair
$I$--$S$ (with $I =$ B or C and $S =$ Cr), the WC-SRO parameter is defined as
\begin{equation}
\alpha_{IS} = 1 - \frac{P_{(S|I)}}{c_S},
\end{equation}
where $P_{(S|I)}$ is the conditional probability of finding a solute atom $S$ as a nearest neighbor to an
interstitial atom $I$, and $c_S$ is the local atomic fraction of solute $S$ within the grain boundary
region. In this convention, $\alpha_{IS} < 0$ indicates preferential ordering (solute enrichment around
the interstitial), $\alpha_{IS} > 0$ indicates avoidance or clustering, and $\alpha_{IS} = 0$ corresponds
to a random solid solution.

To isolate interfacial chemistry, the grain boundary region was defined as a planar slab centered on the interface
plane with a width $w$ equivalent to the derived grain boundary width. Only atoms within
this slab were included in the calculation of the reference solute concentration $c_S$, ensuring that
the SRO metric reflects interfacial rather than bulk chemical statistics.

To capture thermal fluctuations and site-to-site variability at finite temperature, local WC-SRO values
$\{\alpha_{IS,i}\}_{i=1}^{N}$ were computed for every interstitial atom across multiple snapshots extracted
from the final 25~ps of each molecular dynamics trajectory at $T=300$~K.
\section*{Author Contributions}
\textbf{T.D.D} Writing - original draft, writing -- review \& editing, visualization, validation, software, methodology, investigation, formal analysis, data curation. \textbf{R.F.} and \textbf{J. Li} Project administration and supervision, funding acquisition, writing - review \& editing. All authors contributed to the conceptualization of this project.

\section*{Declaration of Competing Interests}
The authors declare that they have no known competing financial interests or personal relationships that could have appeared to influence the work reported in this paper.

\section*{Acknowledgments}
J. Li acknowledges support from National Science Foundation, USA CMMI-1922206 and DMR-1923976.

\section*{Supplemental Materials}
Includes detailed derivations and supporting analyses: 
the protocol for constructing the custom high-angle grain boundary bicrystal system; the hierarchical convergence and error-scaling procedure for determining $\Delta\mu_{\mathrm{Cr}\rightarrow\mathrm{Ni}}$ alongside the composition--temperature dependence and linear regression of $\Delta\mu(x_\mathrm{Cr},T)$; the formulation of interstitial chemical potentials and equilibrium site occupancies: convergence curves; state variable numeric values: spectral analysis results;
and validation benchmarks of the PreFerred Potential against density functional theory.

\section*{Code Availability}
The atomistic simulation routine is available at \url{https://github.com/tylerdolezal/hybrid_MCMD}.

\section*{Data Availability}
Data generated from this work is available at \url{https://github.com/tylerdolezal/gb-interstitial-thermodynamics}.


\bibliographystyle{ieeetr}

\end{document}